# Search for OH 18-cm radio emission from 1I/2017 U1 with the Green Bank telescope

Ryan S. Park[1], D. J. Pisano[2,3,4], T. Joseph W. Lazio[1],
Paul W. Chodas[1], and Shantanu P. Naidu[1]

1. Jet Propulsion Laboratory, California Institute of Technology, Pasadena, California 91109 USA
2. Department of Physics and Astronomy, West Virginia University, Morgantown, WV 26506 USA
3. Gravitational Wave and Cosmology Center, Chestnut Ridge Research Building, Morgantown, WV 26505 USA
4. Adjunct Astronomer at Green Bank Observatory, Green Bank, WV 24944 USA

*ABSTRACT*

This paper reports the first OH 18-cm line observation of the first detected interstellar object 1I/2017 U1 (`Oumuamua) using the Green Bank Telescope. We have observed the OH lines at 1665.402 MHz, 1667.359, and 1720.53 MHz frequencies with a spectral resolution of 357 Hz (approximately 0.06 km-s$^{-1}$). At the time of the observation, `Oumuamua was at topocentric distance and velocity of 1.07 au and 63.4 km-s$^{-1}$, respectively, or at heliocentric distance and velocity of 1.8 au and 39 km-s$^{-1}$, respectively. Based on a detailed data reduction and an analogy-based inversion, our final results confirm the asteroidal origin of `Oumuamua (as discussed in Meech et al., 2017) with an upper bound of OH production of $Q[OH] < 0.17 \times 10^{28}$ s$^{-1}$.

## 1. Introduction

On 2017 October 19, the Pan-STARRS1 asteroid survey telescope system detected a fast-moving small object that turned out to be very different from the hundreds of other similar discoveries it made that year. After just a few days of observation, trajectory computations revealed that the object was following a hyperbolic path relative to the Sun, with an eccentricity of ~1.2 (i.e., clearly greater than 1.0), making this the first detected interstellar object. Initially designated as comet C/2017 U1 (Pan-STARRS), the object was reclassified as an asteroid after a very deep stacked image detected no hint of cometary activity despite its recent close approach to the Sun [MPEC 20176-U183]. Within two weeks, the International Astronomical Union (IAU) assigned an official number and name to the object: 1I/2017 U1 ('Oumuamua).

When 'Oumuamua was discovered, it was already outbound from the Sun, and past its closest approach to the Earth. An extensive observing campaign was quickly organized to characterize the asteroid while it was still relatively near the Earth. Using a suite of optical facilities, including the 3.6-meter Canada-France-Hawaii Telescope (CFHT), the European Southern Observatory's 8-meter Very Large Telescope (VLT) and the 8-meter Gemini South Telescope, Meech et al. (2017) took spectroscopic and high-resolution light-curve measurements. Although the object's spectral characteristics were similar to those seen for objects in our Solar System, the light-curve showed an extreme variation. Ranging over at least 2.5-magnitudes, the light-curve observations indicate that the object has an extremely elongated shape, with an axis ratio of potentially as large as 10:1 with a mean radius of 102 m, and a rotation period of approximately 7 hours (Fraser et al. 2017; Jewitt et al. 2017; Meech et al. 2017). The asteroidal nature of `Oumuamua and its highly elongated shape are two highly surprising results that call into question planetary formation models which suggest that more interstellar objects would be of cometary origin (Meech et al., 2016; Raymond et al., 2017).

Within our own solar system, observations of the 18-cm lines of OH have been a standard method for measuring the water outgassing rate from comets, and more generally for characterizing their volatile contents (e.g., Crovisier et al. 2002, 2016). If volatiles can be discovered in `Oumuamua, it would provide an important constraint on the nature of this body (asteroid vs. comet), and therefore, the dynamics of its original planetary system and the mechanism that could have ejected it. This paper reports the OH 18-cm line observations of `Oumuamua using the Green Bank Telescope (GBT), which is the *first* OH-line observation campaign of an interstellar object.

## 2. Observations and Data Reduction

The observation campaign of `Oumuamua started on UT 2017 November 12, 21:15 for a duration of 4 hours (Project ID: GBT17B-419) using the Green Bank Telescope[1]. During this period, `Oumuamua's rate of change of apparent right ascension[2] and declination were about 26.9"/hr and 7.8"/hr, respectively. We used the VEGAS backend to observe a 23.44 MHz bandwidth in four spectral windows centered on HI at 1420.406 MHz and OH at 1665.402 MHz, 1667.359 MHz, and 1720.53 MHz, with a common spectral resolution of 357 Hz (approximately 0.06 km-s$^{-1}$) for all spectral windows. We conducted standard position-switched observations of 3C48 for flux calibration, assuming an opacity of 0.01. For our observations of `Oumuamua, we used in-band frequency-switching with a throw of ±3 MHz during a series of 5-minute scans with a typical $T_{sys} \cong 17$ K. Our observations were made in the topocentric reference frame while tracking the change in velocity of the target due to the relative motion of the GBT and `Oumuamua.

We reduced the data separately for each spectral window and polarization using standard routines in GBTIDL[3]. We averaged together the calibrated data from all scans and both polarizations while shifting the velocities to align with the first scan. We baselined the resulting spectrum using a first order baseline fit to a 100 km s$^{-1}$ region spanning the target's velocity. With an effective integration time of about 3 hours, the resulting noise is about 3.8 mJy per 0.06 km s$^{-1}$ channel in each spectral window. Finally, we transformed the spectra into the heliocentric reference frame (Figure 1).

---

[1] GBT time was obtained through an agreement between West Virginia University and Green Bank Observatory.
[2] The right ascension rate is multiplied by the cosine of the declination.
[3] http://gbtidl.nrao.edu

At the time of the observation, `Oumuamua was at an approximate heliocentric distance and velocity of 1.8 au and 39 km-s$^{-1}$, respectively. The corresponding topocentric (i.e., `Oumuamua with respect to GBT) distance and topocentric velocity were 1.07 au and 63.4 km-s$^{-1}$, respectively. There are no significant emission or absorption lines seen in any of the OH transitions. We also smoothed the data to 2.4 km-s$^{-1}$ resolution, a more typical velocity width of OH emission and absorption seen in comets (§3). The noise level in the smoothed spectrum was 1.1 mJy; this value is not what would be expected on the basis of radiometer noise alone, which we attribute to residual structure in the estimated spectral baseline. There is no change in the conclusion that no significant OH lines were visible nor did examining the two linear polarizations separately reveal any OH lines. Further, no OH lines were seen in individual 5-minute scans, as may be expected if water was localized on a part of `Oumuamua. While we searched for

**Figure 1.** The OH lines at 1665, 1667, and 1720 MHz spectra for `Oumuamua in a topocentric reference frame (lower x-axis) and a heliocentric reference frame (upper x-axis), presented at the native 0.06 km-s$^{-1}$ velocity resolution. Each transition is offset vertically by 0.02 Jy. At the time of the observation, `Oumuamua was at a topocentric velocity of ~63 km and a heliocentric velocity of ~39 km-s$^{-1}$ (dashed vertical line).

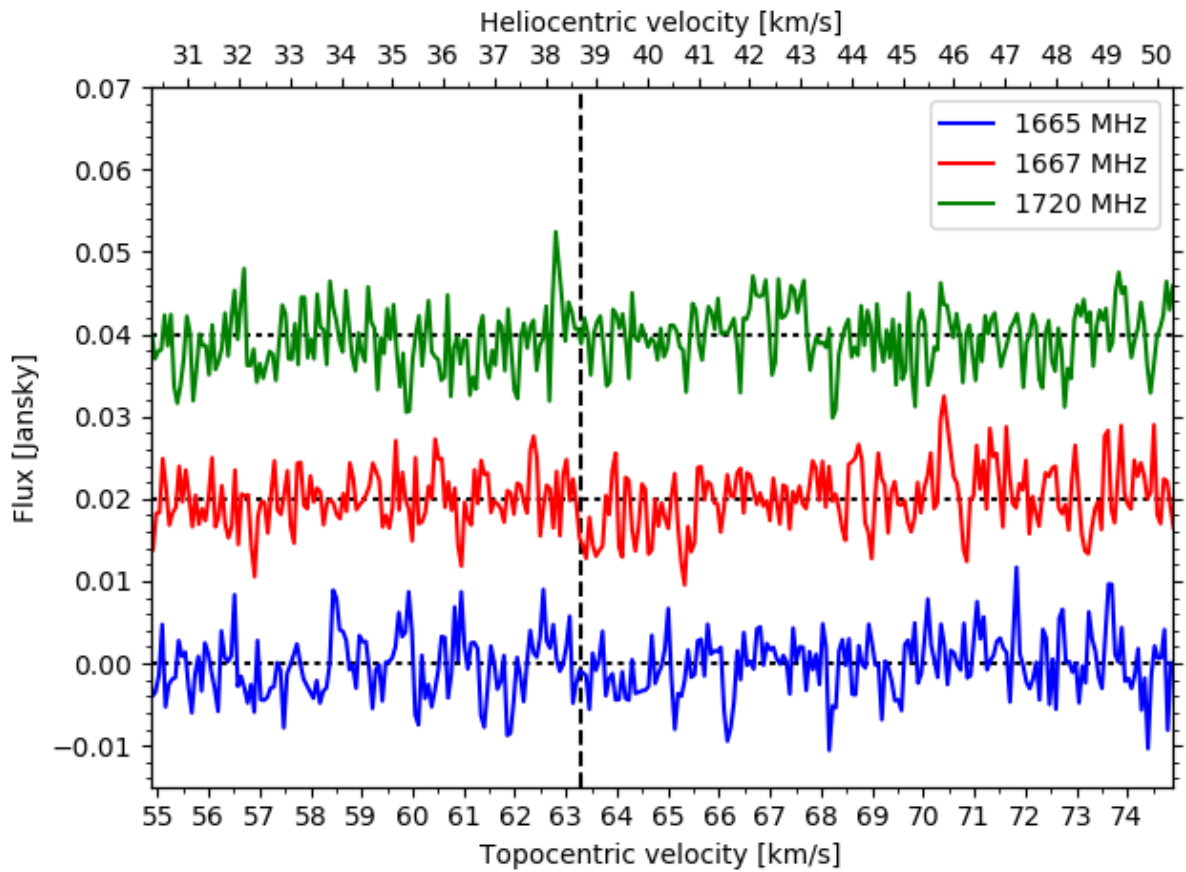

   

HI toward `Oumuamua, its velocity is such that it is on the wing of Milky Way HI emission. Therefore, it was not possible to put a meaningful limit on any HI associated with `Oumuamua.

## 3. Analysis and Conclusion

Both Despois et al. (1981) and Schleicher & A'Hearn (1988) have considered the physics of the 18-cm OH lines from comets. They find that the OH lines can appear either in absorption or emission, depending on the relative population levels in the ground state. In turn, the relative population levels can be predicted in terms of a comet's heliocentric velocity (e.g., Schleicher & A'Hearn 1988, their Figure 9). There are heliocentric velocities at which the models differ, particularly near *crossing points* (i.e., heliocentric velocities at which the predicted line switches from emission to absorption or vice versa). We observed `Oumuamua at a heliocentric velocity such that our results are not (strongly) model dependent upon the modeling of the population of the OH states. Specifically, at the time of our observations, `Oumuamua had a heliocentric velocity just under 40 km-s$^{-1}$, and both Despois et al. (1981) and Schleicher & A'Hearn (1988) make strong predictions that any OH lines should appear in *emission*. Consequently, we conclude that `Oumuamua displays little or no OH production (Figure 1), in qualitative agreement with the lack of a visible coma and consistent with an asteroidal nature.

With no clear emission line apparent in the OH spectra, we adopt the following approach to obtain a semi-quantitative limit on the OH production rate of `Oumuamua. The Nançay database (Crovisier et al. 2002) lists a few comets for which there have been robust detections of OH emission at a time when their heliocentric velocities were comparable to that of `Oumuamua (typically within 2 km-s$^{-1}$) such that the expected ground state inversion factor should be similar; Table 1 summarizes relevant data for these comets. The Nançay database lists two different estimates for the OH production rate Q[OH]. We use the first estimate, which is based on the Despois et al. (1981) inversion and a Haser-equivalent model; differences between the two estimates are minor for this discussion.

Immediately apparent is that our observations were conducted when `Oumuamua was significantly farther away from the Sun (1.8 au) than the previous OH detections (≈ 1 au). However, in terms of distance from the Earth, and therefore potential detectability, `Oumuamua was comparable to or even closer than many of the comets that have been detected. The diversity of the different comets is also apparent, for example with C/1998 J1 (SOHO) having a much higher OH production even though, when observed, it was one of the relatively more distant comets from the Sun (cf. C/1996 B2 Hyakutake at 0.56 au from the Sun).

In order to obtain (semi-)quantitative limits, we require an estimate for the "area" of the OH line (mJy km s$^{-1}$). As noted in §2, the noise level in a 2.4 km s$^{-1}$ spectral bin is 1.1 mJy. We assume a statistical weight of OH(1665 MHz):OH(1667 MHz) = 9:5 (Despois et al. 1981), and adopt 3σ as an upper limit. The resulting area is $A$ = 2.9 mJy km s$^{-1}$ (see Table 1).

We produce two, admittedly crude, limits on the OH production rate $Q$[OH] of `Oumuamua. First, we note that Comet C/1998 J1 (SOHO) was observed at a similar distance from the Earth (i.e., 1.05 au), but when it was approximately two times closer to the Sun (i.e., 0.96 au vs. 1.80 au). The OH line emission is generated by OH ions, which are themselves produced by photodissociation of a parent molecule

    

(likely water). The production of this parent molecule should scale with the solar UV illumination (Despois et al. 1981), but this scaling is not the simple $d_\odot^{-2}$ for a comet-Sun distance $d_\odot$. Rather, the scaling is as $d_\odot^{1/4}$, as a comet is assumed to have a gray body atmosphere in which the parent molecule expands before photoionization. We scale the OH production rate by a similar factor. Also taking into account the ratio between the "areas" for the two objects (262 mJy km s$^{-1}$ vs. 2.9 mJy km s$^{-1}$), the scaling equation becomes:

$$Q_{\text{`Oumuamua}} = Q_{\text{Comet}} \frac{A_{\text{`Oumuamua}}}{A_{\text{Comet}}} \left( \frac{d_{\odot\text{Comet}}}{d_{\odot\text{`Oumuamua}}} \right)^{1/4} \qquad (1)$$

where $A$ is the area of the line for each object. Equation (1) assumes that, if `Oumuamua were producing OH, then its properties would be similar to those of a solar system comet. The interested reader is referred to Despois et al. (1981, Part II) for additional details underlying this approximation. We find an upper limit on OH production of $Q[\text{OH}] < 0.17 \times 10^{28}$ s$^{-1}$, which is the first estimate recorded for `Oumuamua in Table 1.

For the second estimate, we also take into account that, for the other comets, the strength of the emission line would scale as $d_\oplus^{-2}$ for the comet-Earth distance $d_\oplus$, under the assumption that any putative coma of `Oumuamua is less than the size of the GBT beam (~ 6', equivalent to 0.002 au) at its distance at the time of the observation, or including an additional factor of $\left( d_{\oplus\text{`Oumuamua}}/d_{\oplus\text{Comet}} \right)^2$ in Eqn. (1). We find an upper limit on the OH production of $Q[\text{OH}] < 0.15 \times 10^{28}$ s$^{-1}$, which is the second estimate recorded (see Table 1). Since the first estimate is more conservative, we report $Q[\text{OH}] < 0.17 \times 10^{28}$ s$^{-1}$ as the upper limit in the abstract.

Although not listed in Table 1 the Nançay database also reports the OH lines of comets 1P/Halley and C/1995 O1 (Hale-Bopp) at various heliocentric distances, including 1.6-2.0 au (i.e., similar to the distance when `Oumuamua was observed), when they were moving away from the Sun. The OH production rates for these large comets (i.e., 11 km for 1P/Halley and 60 km for Hale-Bopp) were generally well above 10 × 10$^{28}$ s$^{-1}$.

We are unaware of any detections or reported upper limits on the OH-18 cm emission from asteroids (e.g., Rivkin et al. 2015). The absence of such emission is in qualitative agreement with our results. Also, we note that we have not made a distinction in the analysis whether `Oumuamua is on an inbound trajectory vs. outbound trajectory, which may change the OH emission slightly (e.g., an outbound comet should have been heated more and may show different OH emission).

While no OH lines were detected from `Oumuamua, as Jewitt et al. (2017) discuss, the absence of cometary properties does not imply the absence of volatiles. Indeed, as they note, it is quite possible that `Oumuamua may become active even as it recedes from the Sun as the thermal pulse that occurred during perihelion diffuses into its interior. Should that be the case, further OH line observations would be warranted. We note that a similar GBT survey was conducted about month after our observation at a topocentric distance of ~2 au and no cometary activity was observed as well (Enriquez, 2018).

Finally, as the first interstellar object, whether the lack of activity displayed by `Oumuamua is typical cannot be assessed. The properties and trajectories of other interstellar objects discovered in the future

    

may be sufficiently different that at least some of them will display cometary activity, for which OH line observations would be warranted as part of the activities to characterize those objects. Moreover, even if `Oumuamua was asteroidal, there are predictions of cometary bodies also being ejected and future OH observations of interstellar objects in the solar system would be warranted (Jackson et al., 2017).

Table 1. Comets with detectable OH emission and heliocentric velocities comparable to 1I/2017 U1. Note that data for all objects are taken from the Nançay comet database except for 1I/2017 U1. For 1I/2017 U1, see discussion in the text. Note that the OH spectra at an offset position are all within the GBT beam width.

| Object | Distance from Earth (au) | Distance from Sun (au) | Area, $A$ (mJy km s$^{-1}$) | $Q$[OH] ($\times 10^{28}$ s$^{-1}$) |
|---|---|---|---|---|
| 96P/Machholz 1 | 0.72 | 0.84 | 74 | 2.9 |
| C/2004 F4 (Bradfield) | 1.42 | 0.91 | 61 | 4.5 |
| C/2002 X5 (Kudo-Fujikawa) | 0.90 | 0.92 | 80 | 3.4 |
| C/2002 V1 (NEAT) | 1.67 | 1.05 | 65 | 7.1 |
| C/1998 J1 (SOHO) | 1.05 | 0.96 | 262 | 17.6 |
| C/1996 B2 (Hyakutake) | 1.20 | 0.56 | 90 | 11.7 |
| Machholz 1988 V | 1.35 | 0.90 | 48 | 1.7 |
| 1I/2017 U1, first estimate | 1.07 | 1.80 | <2.9 | <0.17 |
| 1I/2017 U1, second estimate | 1.07 | 1.80 | <2.9 | <0.15 |

**Acknowledgements**
We thank the developers and maintainers of the Nançay comet database, without whom this work would have been far more difficult, and the referee who made several comments that clarified both the analysis and the presentation. This research has made use of NASA's Astrophysics Data System. We also thank the West Virginia University Research Office for its support of the operations of the Green Bank Telescope. GBT is a facility of the National Science Foundation operated under cooperative agreement by Associated Universities, Inc. Part of this research was carried out at the Jet Propulsion Laboratory, California Institute of Technology, under a contract with the National Aeronautics and Space Administration.